\documentclass[final,3p,times,twocolumn]{elsarticle}
\usepackage{graphics}
\usepackage{epsfig}

\usepackage{amssymb}

\journal{Physics Letter B}

\begin{document}

\begin{frontmatter}



\title{Magic nuclei in superheavy valley}


\author{M. Bhuyan}

\address{School of Physics, Sambalpur University, Jyotivihar,
Burla-768019, INDIA.}

\author{S. K. Patra}

\address{Institute of Physics, Sachivalaya Marg, Bhubaneswar-751005,
INDIA.}

\begin{abstract}
An extensive theoretical search for the proton magic number in the 
superheavy valley beyond $Z=$82 and corresponding neutron magic number 
after $N=$126 is carried out. For this we scanned a wide range of 
elements $Z=112-130$ and their isotopes. The well established 
non-relativistic Skryme-Hartree-Fock and Relativistic Mean Field 
formalisms with various force parameters are used. Based on the 
calculated systematics of pairing gap, two neutron separation energy 
and the shell correction energy for these nuclei, we find $Z=$120 
as the next proton magic and N=172, 182/184, 208 and 258 the subsequent 
neutron magic numbers.
\end{abstract}


\end{frontmatter}

After the discovery of artificial transmutation of elements by Sir
Ernest Rutherford in 1919 \cite{ruth19}, the search for new elements is
an important issue in nuclear science. The existence of
elements beyond the last heaviest naturally occurring $^{238}$U,
i.e., the  discovery of Neptunium, Plutonium and other 14 elements
(transuranium elements), which make a separate block in Mendeleev$'$s
periodic table was a revolution in the Nuclear Chemistry.
This enhancement in the periodic table raises a few questions in our mind:

\begin{itemize}
\item Whether there is a limited number of elements that can co-exist
either in nature or can be produced from artificial synthesis by using
modern technique ?
\item What is the maximum number of protons and neutrons that of a nucleus ?
\item What is the next double shell closure nucleus beyond $^{208}$Pb ?
\end{itemize}

To answer these questions, first we have to understand the agent which is
responsible to rescue the nucleus against Coulomb repulsion. The obvious
reply is the shell energy, which stabilises the nucleus against
Coulomb disintegration \cite{satpathy04}. Many theoretical models,
like the macroscopic$-$microscopic (MM) calculations to
explain involve some prior knowledge of
densities, single-particle potentials and other bulk properties which may
accumulate serious error in the largely extrapolated mass region of interest.
They predict the magic shells at $Z$=114 and $N$=184
\cite{patyk91,nix94,mosel69,nilsson69} which could have surprisingly
long life time even of the order of a  million years
\cite{myers65,sobi66,meldner67,mosel69,buro05}.  Some other such predictions of
shell-closure for the superheavy region within the relativistic and
non-relativistic theories depend mostly on the force parameters
\cite{rutz97,rutz99}.

Experimentally, till now, the quest for superheavy nuclei
has been dramatically rejuvenated in recent years owing to the
emergence of hot and cold fusion reactions. In cold fusion reactions
involving a doubly magic spherical target and a deformed projectile
were used by GSI \cite{hof00,hof95,hoff95,hof96,hof98} to produce
heavy elements upto $Z$ = 110$-$112. In hot fusion evaporation
reactions with a deformed transuranium target and a doubly magic
spherical projectile were used in the synthesis of superheavy nuclei
$Z$ = 112$-$118 at Dubna \cite{oga98,oga01,oga04,oga07,oga10,oga11,eic07}.
At the production time of $Z$ = 112 nucleus at GSI the fusion cross
section was extremely small ($1 pb$), which led to the conclusion
that reaching still heavier elements will be very difficult. At this
time, the emergence of hot fusion reactions using $^{48}Ca$ projectiles
at Dubna has dramatically changed the situation and nuclei with
$Z$ = 114$-$118 were synthesized and also observed their $\alpha$-decay
as well as terminating spontaneous fission events. It is observed
that $Z$ = 115$-$117 nuclei have long $\alpha$-decay chains contrast to
the short chains of $Z$ = 114$-$118. Moreover, the life times
of the superheavy nuclei with $Z$ = 110$-$112 are in milliseconds
and microseconds whereas the life time of $Z$ = 114$-$118 up to $30$ s. 
This pronounced increase in life times for these heavier nuclei has 
provided great encouragement to search the magic number somewhere beyond 
$Z=$114. Moreover, it is also an interesting and important question for 
the recent experimental discovery \cite{marinov07,marinov09,bhu09}
say chemical method of $Z$ = 122 from the natural $^{211,213,217,218}$Th 
which have long lived superdeformed (SD) and/ or hyperdeformed (HD) isomeric  
states 16 to 22 orders of magnitude longer than their corresponding 
ground-state (half-life of $^{292}$122 is $t_{1/2}\geq 10^8$ years).

\begin{figure}[ht]
\vspace{0.65cm}
\begin{center}
\includegraphics[width=1.0\columnwidth,height=1.25\columnwidth]{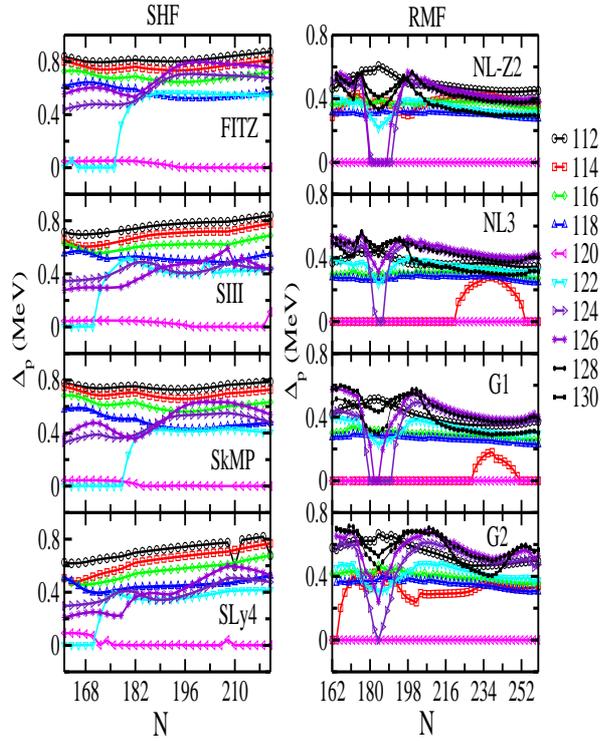}
\caption{The proton average pairing gap $\Delta_p$ for Z=112-126 with
N=162-220 and Z=112-130 with N=162-260.
}
\end{center}
\label{Fig. 1}
\end{figure}

\begin{figure}[ht]
\vspace{0.65cm}
\begin{center}
\includegraphics[width=1.0\columnwidth,height=1.25\columnwidth]{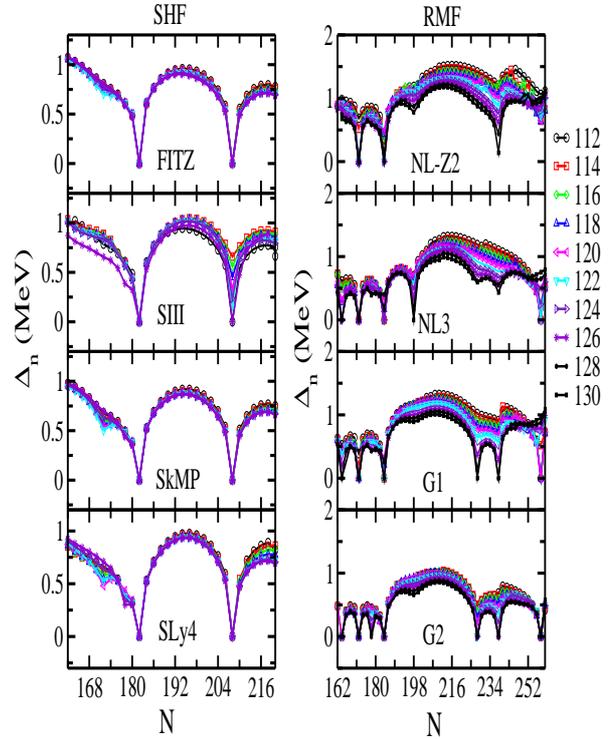}
\caption{Same as FIG.1 but for neutron average pairing gap $\Delta_n$.
}
\end{center}
\label{Fig. 2}
\end{figure}

\begin{figure}[ht]
\vspace{0.65cm}
\begin{center}
\includegraphics[width=1.0\columnwidth,height=1.25\columnwidth]{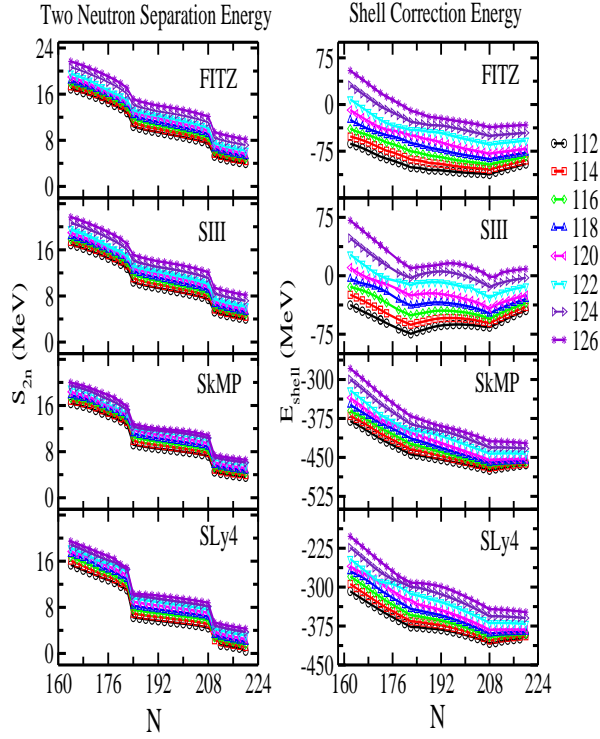}
\caption{The two neutron separation energy $S_{2n}$ and the shell
correction energy $E_{shell}$ for $Z$=112-126 and $N$=162-220
in the framework of SHF theory
}
\end{center}
\label{Fig. 3}
\end{figure}

In this letter, our aim is to look for the next double
closed nucleus beyond $^{208}$Pb which may be a possible
candidate for the experimentalists to look for. For this, we have
used two well-defined but distinct approaches (i) non-relativistic
Skryme-Hartree-Fock (SHF) with FITZ, SIII, SkMP and SLy4 interactions
\cite{cha97,stone07}
(ii) Relativistic Mean Field (RMF) formalism \cite{sero86,ring90}
with NL3, G1, G2 and NL-Z2 parameter sets. These models have been
successfully applied in the description of nuclear structure
phenomena both in $\beta-$stable and $\beta-$unstable regions
throughout the periodic chart. The constant strength scheme
is adopted to take care of pairing correlation \cite{estal01}
and evaluated the pairing gaps $\triangle_n$ and $\triangle_p$ for
neutron and proton respectively from the celebrity BCS equations
\cite{preston}.

We scanned a wide range of nuclei starting from the
proton-rich to the neutron-rich region in the superheavy valley
(Z=112 to Z=130). It is well understood and settled that the
properties of a magic number for a nuclear system has the
following characteristics:

\begin{itemize}
\item The average pairing gap for proton $\Delta_p$ and neutron
$\Delta_n$ at the magic number is minimum.
\item The binding energy per particle is maximum compared to the neighboring
one, i.e.there must be a sudden decrease (jump) in two neutron
(or two proton) separation energy $S_{2n}$ just after the magic number in an
isotopic or isotonic chain.
\item At the magic number, the shell correction energy $E_{shell}$ is
maximum negative. In other words, a pronounced energy gap in the
single-particle levels $\epsilon_{n,p}$ appears at the magic number.
\end{itemize}

We focus on the shell closure properties in the superheavy valley
based on the above three important observables and identify the magic
proton and neutron numbers.

\begin{table}
\caption{Single-particle levels for neutron $\epsilon_{n}$ (MeV) for
$^{302}$120 in SHF(SLy4 and FITZ) and $^{304}$120 in  RMF (NL3 and G2).
}
\renewcommand{\tabcolsep}{0.35cm}
\renewcommand{\arraystretch}{1.0}
\begin{tabular}{|c|c|c|c|c|c|c|c|c|c|c|c|c|c|c|c|c|}
\hline
Orbit&\multicolumn{4}{|c|}{neutron ($\epsilon_p$)} \\
\hline
& SLy4& FITZ&  NL3& G2 \\
\hline
$s^{1/2}$&-38.6&-34.6&-39.8&-38.8\\
$p^{3/2}$&-34.8&-31.1&-36.3&-35.1\\
$p^{1/2}$&-34.6&-31.0&-36.1&-34.8\\
$d^{5/2}$&-29.9&-26.6&-31.4&-30.2\\
$d^{3/2}$&-29.2&-26.1&30.7&-29.3\\
$s^{1/2}$&-26.2&-23.1&-26.3&-26.1\\
$f^{7/2}$&-24.2&-21.3&-25.7&-24.5\\
$f^{5/2}$&-22.7&-20.2&-24.2&-22.8\\
$p^{3/2}$&-19.1&-16.5&-19.8&-19.0\\
$p^{1/2}$&-18.9&-16.3&-19.7&-18.7\\
$g^{9/2}$&-17.9&-15.3&-19.3&-18.1\\
$g^{7/2}$&-15.3&-13.4&-17&-15.4\\
$d^{5/2}$&-11.9&-9.5&-12.9&-11.6\\
$h^{11/2}$&-11.1&-8.8&-12.5&-11.3\\
$d^{3/2}$&-10.9&-8.7&-12.3&-10.7\\
$s^{1/2}$&-9.8&-7.2&-10.2&-9.3\\
$h^{9/2}$&-7.3&-6.0&-9.3&-7.5\\
$f^{7/2}$&-4.5&-2.4&-5.8&-4.1\\
$i^{13/2}$&-4.0&-2.0&-5.5&-4.1\\
$f^{5/2}$&-2.6&-0.9&-4.7&-2.4\\
$p^{3/2}$&-1.4&0.4&-2.6&-0.8\\
\hline
\end{tabular}
\label{Table 1}
\end{table}

The average pairing gap for proton $\Delta_p$ and for neutron $\Delta_n$
are the representative of strength of the pairing correlations. The curves
for $\Delta_p$ are displayed in FIG. 1 obtained by SHF and RMF with
FITZ, SIII, SLy4, SkMP and NL3,NL-Z2, G1, G2 force
parameterizations. If we investigate the figure carefully,
it is clear that the value of $\Delta_p$ almost zero for
the whole Z=120 isotopic chain in both the theorical approaches.
A similar $\Delta_p$ is observed for few cases of Z=124 and Z=114
isotopes.

To predict the corresponding neutron shell closure of the magic
$Z$=120, we have estimated the neutron pairing gap $\Delta_n$ for
all elements $Z$=112$-$130 with their corresponding isotopic chain.
As a result of this, the calculated $\Delta_n$ for the whole atomic
nuclei in the isotopic chains are displayed in FIG. 2. We obtained
an arc like structure with vanishing $\Delta_n$ at
$N$= 182, 208 and $N$=172, 184, 258 respectively for SHF and RMF
of the considered parameter sets. Further, the neutron pairing gap
is found to be minimum among the isotopic chains pointing towards
the magic nature of $Z=$120.  Therefore, all of these force parameters
are directing $Z=120$ as the next magic number after $Z=$82.

\begin{table}
\caption{Same as Table 1. but for neutron $\epsilon_{n}$ (MeV).
}
\renewcommand{\tabcolsep}{0.35cm}
\renewcommand{\arraystretch}{1.5}
\begin{tabular}{|c|c|c|c|c|c|c|c|c|c|c|c|c|c|c|c|c|}
\hline
orbit& \multicolumn{4}{|c|}{proton ($\epsilon_p$)} \\
\hline
&SLy4&  FITZ&  NL3&  G2 \\
\hline
$s_{1/2}$ & -58.0 & -50.7&-55.4&-54.0\\
$p_{3/2}$ & -53.7 & -47.4&-51.8&-50.2\\
$p_{1/2}$ & -53.4 & -47.2&-51.6&-50.0\\
$d_{5/2}$ & -48.0 & -42.9&-46.7&-45.1\\
$d_{3/2}$ & -47.2 & -42.3&-46.0&-44.3\\
$s_{1/2}$ & -43.8 & -39.2&-41.0&-40.3\\
$f_{7/2}$ & -41.5 & -37.5&-40.6&-39.1\\
$f_{5/2}$ & -39.9 & -36.4&-39.3&-37.6\\
$p_{3/2}$ & -36.0 & -32.8&-34.6&-33.5\\
$p_{1/2}$ & -35.8 & -32.5&-34.5&-33.1\\
$g_{9/2}$ & -34.2 & -31.5&-33.9&-32.5\\
$g_{7/2}$ & -31.7 & -29.6&-31.8&-30.0\\
$d_{5/2}$ & -28.0 & -26.2&-27.8&-26.3\\
$d_{3/2}$ & -26.8 & -25.2&-27.2&-25.4\\
$h_{11/2}$& -26.5 & -25.0&-26.9&-25.3\\
$s_{1/2}$ & -25.1 & -24.1&-24.8&-23.3\\
$h_{9/2}$ & -22.7 & -22.2&-23.8&-21.8\\
$f_{7/2}$ & -19.8 & -19.2&-20.5&-18.7\\
$i_{13/2}$& -18.5 & -18.1&-19.6&-18.1\\
$f_{5/2}$ & -17.7 & -17.5&-19.4&-16.9\\
$p_{3/2}$ & -16.5 & -15.9&-16.9&-14.9\\
$p_{1/2}$ & -16.2 & -15.7&-16.7&-14.4\\
$i_{11/2}$& -13.3 & -14.1&-15.6&-13.1\\
$g_{9/2}$ & -11.7 & -11.9&-13.2&-11.0\\
$j_{15/2}$& -10.3 & -10.9&-12.1&-10.5\\
$g_{7/2}$ & -8.8  & -9.6&-11.5& -8.6 \\
$d_{5/2}$ & -8.0  & -8.5&-9.5&  -7.2 \\
$d_{3/2}$ & -7.0  & -7.7&-9.2&  -6.6 \\
$s_{1/2}$ &-3.6   & -5.7&-8.2&  -6.0 \\
$j_{13/2}$&       & &-7.3&      -4.3 \\
\hline
\end{tabular}
\label{Table 1}
\end{table}

As mentioned earlier, the binding energy per particle (BE/A)
is maximum for double closed nucleus compared to the neighbouring one.
For example, the BE/A with SHF (FITZ set)
for $^{300,302,304}120$ are 7.046, 7.048 and 7.044 MeV corresponding
to N=180, 182 and 184 respectively. Similarly with SLy4 these values
are 6.950, 6.952 and 6.933 MeV.
This is reflected in the sudden jump of $S_{2n}$ from a higher
value to a lower one at the magic number in an isotopic chain.  This
lowering in two neutron separation energy is an acid test for shell
closure investigation. FIG. 3 shows the $S_{2n}$ as a function of
neutron number for all the isotopic chain of the considered elements
for both SHF and RMF formalisms. In spite of the complexity about
single-particle and collective properties  of the nuclear
interaction some simple phenomenological facts emerge from the bulk
properties of the low-lying states in the even-even atomic
nuclei.
The $S_{2n}$ energy is sensitive to this collective/single-particle
inter play and provides sufficient information about the nuclear
structure effects. From FIG. 3, we notice such effect, i.e., jump in
two neutron separation energy at N=182 and 208 with SHF. However, such
jumps are found at N=172, 184, 258 in RMF calculations confirming the
shell closure properties of the neutron numbers.

The shell correction energy $E_{shell}$ is a key quantity
to determine the shell closure of nucleon. This concept
was introduced by Strutinski \cite{stru67} in liquid-drop model
to take care of the shell effects. As a result, the
whole scenario of liquid properties converted to shell structure
which could explain the magic shell even in the frame-work of
liquid-drop model. The magnitude of total (proton plus neutron)
$E_{shell}$ energy is dictated by the level density around the
Fermi level. A positive $E_{shell}$ reduces the binding energy
and a negative shell correction energy increases the stability
of the nucleus. As a representative case, we have depicted
our SHF result of $E_{shell}$ in FIG. 3. It is clear from the
figure the extra stability of $^{302,328}$120. We find similar
results of large negative shell energy for RMF calculation
at neutron number 172, 184, 258. Such calculations for few
cases are reported in Ref. \cite{sil04}.

As a further confirmative test, the single-particle energy levels
for neutrons and protons $\epsilon_{n,p}$ are analyzed. The calculated
$\epsilon_{n,p}$ are
presented in Table I for $^{302}$120 SHF(SLy4 and FITZ) and for $^{304}$120
RMF(NL3 and G2) as representative cases. From the Table, one can estimate
the energy gaps $\triangle\epsilon_{n,p}$ for neutron and proton
orbits. For example,  in $^{302}$120 (FITZ), the gap
$\triangle\epsilon_n=\epsilon_n (3d_{3/2})-\epsilon_n(4s_{1/2})$ at
N=182 is $1.977$ MeV which
is comparable with 1.898 MeV of the known magic gap at N=50 for the same
nucleus. For the proton case at Z=120, we get $\triangle\epsilon_p=\epsilon_p
(2f_{5/2})-\epsilon_p(3p_{3/2})=1.340$ MeV which is again a large
gap of similar order ($\triangle\epsilon_p=\epsilon_p
(1g_{9/2})-\epsilon_p(1g_{7/2})=1.862$ MeV) for N=50. Almost
identical behaviour is noticed with other SHF and RMF (at N=184) calculations,
irrespective of parameter used, confirming Z=120 as a clear magic number.

In summary, we have analyzed the pairing gap $\Delta_p$ and
$\Delta_n$, two-neutron separation energy $S_{2n}$, shell
correction energy $E_{shell}$ and single-particle energy
for the whole $Z=$112$-$130 region
covering the proton-rich to neutron-rich isotopes. To our knowledge,
this is one of the first such extensive and rigorous calculation
in both SHF and RMF models using a large number of parameter sets. The
recently developed effective field theory motivated relativistic
mean field forces G1 and G2 are also involved. Although the results
depend slightly on the forces used, the general set of magic numbers
beyond $^{208}$Pb are Z=120 and N=172, 182/184, 208 and 258. The
highly discussed proton magic number $Z=114$ in the past (last four
decades) is found to be feebly magic in nature.

We thank Profs. L. Satpathy, C.R. Praharaj and K. Kundu for discussions and
a careful reading of the manuscript. One of the authors (MB) thank Institute
of Physics for hospitality.

\end{document}